\documentstyle[twocolumn,aps]{revtex}
\begin{document}
\draft
\title{ A Local Realistic Model for
 Two-Particle Einstein-Podolsky-Rosen Pairs }
\author{ Zhi-Yuan Li } 
\address{Ames Laboratory, US Department of Energy and
Department of Physics and Astronomy, Iowa State University, 
Ames, Iowa 50011, and
Institute of Physics, Chinese Academy of Sciences, P. O. Box 603,
Beijing 100080, China}
\date{Received~~~~~~~~~June  2001}
\maketitle

\begin{abstract}
 A local realistic model for quantum mechanics of two-particle
Einstein-Podolsky-Rosen pairs is proposed. In this model, it is
the strict obedience of  conservation laws  in each  
event at the quantum level that uphold the perfect correlation
of two spatially-separated particles, instead of nonlocality in
the orthodox formulation of quantum mechanics. Therefore, 
one can conclude that all components of the spin of two particles, and 
the position and momentum of a particle can 
be measured simultaneously. The proposed model yields
the same  statistical prediction on an
ensemble of  individual particles as the orthodox formulation 
does. This suggests that 
the wave  function is not a complete description of individual particle
as assumed in the orthodox  formulation, 
but only a statistical description of an ensemble of  particles.

\end{abstract}
\pacs{ PACS numbers:  03.65.-w, 03.65.Ud, 03.65.Ta }
\narrowtext
\par

The 20th century saw the birth, growth, and prosperity of quantum
mechanics, and its influence in every aspect of science and 
technology.      The prediction of
 quantum theory has been found in excellent agreement with experimental
results in an extremely wide range.
 However, the interpretation of quantum mechanics has
been widely disputed over a long time [1-16], notably coming from 
Einstein in one side and Bohr  in the other.
The central point of the controversy is whether or not
the  wave function is a complete description of an individual system
at the quantum level. In a historic paper [1], Einstein, Podolsky, and 
Rosen (EPR) analyzed   a system consisting of two spatially
separated but quantum-mechanically correlated particles, the so-called
EPR pairs or entangled pairs [3], and 
 argued that both particles could simultaneously have 
predetermined values of non-commuting operators, such as position and 
momentum. Thus,  they concluded that quantum mechanics is
not complete.  In regard to this  kind of entangled system, 
Bell proved an inequality [11] based on which 
it is possible to test  quantum mechanics and the
opponent deterministic hidden-variable 
theory [9-14] in a quantitative manner. 
Most experiments to date have favored quantum mechanics [12,17-20]. 
Today, it has been well-established  in the orthodox 
formulation of quantum mechanics that  spooky 
nonlocal characteristics are indeed present in these EPR pairs.
This, together with the abandonment of realism (enforced in   
 the wave-packet reduction hypothesis in  
old quantum measurement problem) represents the central  
viewpoint of orthodox quantum mechanics on  physics in
microscopic world. Both viewpoints will drastically 
change our concepts and philosophy of nature.

\par
Is it possible to avoid such a radical 
revolution of  concepts about nature by using some
 local realistic viewpoints to understand microscopic world 
while pertaining agreement 
with experimental observations? In this paper
 we will give a positive answer to this question. We will propose
a local realistic model which  
directly comes from  standard  quantum mechanics for many-particle 
systems. Yet, the proposed model starts from a different viewpoint 
and finally leads to  different results from the orthodox formulation. 

\par
To this end, we first consider in general 
the motion of  two  independent  particles under the law of 
quantum mechanics. One typical case is that these particles 
are  separated in space, and noninteracting  with each other.
  The Schr\"odinger equation for
this two-particle system is  written as
$$ i\hbar \frac{\partial}{\partial t}
 \Psi = H\Psi= (H_1+H_2) \Psi. \eqno(1) $$
$H_1({\bf r}_1)$ and $H_2({\bf r}_2)$ are  the Hamiltonian 
for particles 1 and 2, respectively. 
From physical  intuition and simple argument one can find that
 the system wave function $\Psi$ is just
 the  direct  product of the wave function  
of each particle, namely
$$\Psi({\bf r}_1,{\bf r}_2)=\psi_1({\bf r}_1) \psi_2({\bf r}_2). \eqno(2)$$
$\psi_1$ and $\psi_2$ satisfy 
$$i\hbar \frac{\partial}{\partial t}
 \psi_1 =  H_1 \psi_1;~~~~~~~i\hbar \frac{\partial}{\partial t}
 \psi_2 =  H_2 \psi_2.  \eqno(3) $$
The general solution of Eq. (3) is
$$  \psi_1=\sum_{\lambda_1} c_{1,\lambda_1}
 \psi_1({\bf r}_1,\lambda_1),  ~~~~ 
\psi_2=\sum_{\lambda_2} c_{2,\lambda_2}
 \psi_2({\bf r}_2,\lambda_2),   \eqno(4)$$
where $\psi_1({\bf r}_1,\lambda_1)$ and $\psi_2({\bf r}_2,\lambda_2)$ are
 eigenstates of particles 1 and 2 under Hamiltonian $H_1$ and $H_2$,
 respectively,  with $\lambda_1$
($\lambda_2$) being their eigenvalues, and
 $c_1$ and $c_2$ are the superposition coefficients. 
Eqs. (2)-(4) verify that there is no mechanical correlation
between particles 1 and 2, as is exactly what we have supposed that
these two particles are independent.
Any external influence on one particle will not affect the wave
function of the other particle.

\par
The above analysis is also applicable to a special two-particle system,
the free EPR pair, since what we only require in Eqs. (1)-(4)
is that two particles are far-away separated and
non-interacting, no matter these two particles are 
 either  uncorrelated or correlated. 
 To be more specific,
we consider Bohm's scheme with  a pair of spatially separated spin-1/2 
atoms,   which is produced from the dissociation of
a molecule in the spin singlet (total spin 0) 
state (e.g., $\rm Hg_2$ molecule)[5,6]. 
Up to some time $t=0$, the wave function
of this combined system is 
$$\Psi=\frac{1}{\sqrt{2}} [\psi_{+}(1) \psi_{-}(2)-\psi_{-}(1) \psi_{+}(2)],
\eqno(5)$$
 where $\psi_{+}(1)$ means that atom 1 has spin $+\hbar /2$, and so on. 
As $\Psi$ is rotationally invariant, the spin axis ``+" can be selected
 along any  direction  in space. 
This antisymmetric spin wave function is very common in
molecular bound  states. 
A good example is the spin singlet state of a 
hydrogenic molecule, which is formed from
 the group motion of two electrons under nuclear attraction.
It should be  emphasized that two free hydrogenic atoms can 
not form a spin singlet state.

\par
What happens if the two atoms are separated
by a method that does not influence the total spin?  
In the orthodox formulation, the wave function in Eq. (5) 
keeps unchanged even when the two atoms have separated so far away that 
they cease to interact, namely, atoms in 
both the bound state and free state have the same spin-singlet state.
And it is the nonlocality (something of action-at-a-distance) that  
uphold the perfect correlation of the spins of the two atoms. Once
one measures one spin and finds it is along $\bf \hat n$ direction, 
the other spin is immediately known
to  be along  $-\bf \hat n$ even if no measurement is
 performed. Although spooky, this assumption has found   
strong experimental supports for a long time.  
These experiments are closely related to 
the test of Bell's inequality, and  found to favor
 quantum mechanics [10,15-18].
The joint  outcome of  the spin of particle 1   
along the direction $ {\bf n}_1$ and that of  particle 2  
along direction $ {\bf n}_2$ in orthodox quantum mechanics is [12] 
$$E^{\Psi}( {\bf n}_1, {\bf n}_2)=<\Psi|({\bf \sigma}_1 \cdot
 {\bf n}_1) ~~({\bf \sigma}_2 \cdot  {\bf n}_2)|\Psi>
=- {\bf n}_1
\cdot  {\bf n}_2, \eqno(6)$$
where $\sigma$ is the vector of  Pauli matrices. 
The result in Eq. (6)  clearly violates Bell's inequality [11-14].
It is such kinds of  violation that  validate orthodox quantum mechanics 
and  refutes any hidden-variable theory.

\par
In our model, the system wave function 
$\Psi$ has the form  of Eqs. (2) and (4), each atom having its
own wave function independent of the other. Then,  because
the total spin is conserved in every process of molecular dissociation,
the two particles must have a specific spin vector antiparallel to
each other, and thus  both are in a pure spin state each time. 
This  yields to 
$$\Psi=\psi_1 \psi_2, \eqno(7a) $$
$$\psi_1= \psi_{+}(1, {\bf n}),~~~~~~ 
\psi_2= \psi_{-}(2, {\bf n}),  \eqno(7b)$$
where $ {\bf n}$ is a random unit vector of direction  
uniformly distributed in space, in accordance with the rotational invariance
of original spin-singlet state in the molecule.  
The  wave function Eq. (7) changes its form  under axis 
rotation if it refers to  a single spin pair. However, 
since here  it actually represents the whole ensemble 
of spin pairs,  it is rotationally invariant, 
as is clear from the fact that $\bf n$
is uniformly distributed in space. This ensemble is composed of many
spin pairs each of which is composed of two spins
with definite but antiparallel directions.

\par
The physical picture involved in 
Eq. (7) is quite simple. If we assume the spin-singlet molecule as
an atom source, then this source emits  at each time  a pair of
atoms who have definite  spin vector antiparallel to each other, but what
direction the spins are along is unknown.  This is similar
to spontaneous emission of photon from excited atoms. Each time the emitted
photon has a definite polarization, but the direction is unspecified, 
all  are possible. One can  utilize  this similarity to calculate
 the quantum-mechanical expectation value (statistical average) of 
all components of each spin, $\bar S_x$, $\bar S_y$, and so on, 
which are of course all zero, the same as predicted
by the orthodox  formulation. 
The joint outcome  can be found according to Fig. 1, which is  
$$E( {\bf n}_1, {\bf n}_2)= <\psi_1( {\bf n}_1)|
{\bf \sigma}_1 \cdot  {\bf n}_1  |\psi_1( {\bf n}_1)> $$ $$
\times  ~~<\psi_2(- {\bf n}_1) |{\bf \sigma}_2 \cdot 
 {\bf n}_2|\psi_2(- {\bf n}_1) >=- {\bf n}_1
\cdot  {\bf n}_2, \eqno(8)$$
also exactly the same as that predicted by the orthodox formulation of 
quantum mechanics [shown in  Eq. (6)]. Therefore our proposed model
 also passes the test of Bell's inequalities. 
Derivation of Eq. (8) implies
the application  of Malus's law in classical optics. 
It is well-known that when natural light passes through two
polarizers with an inclination of $\theta$, we observe 
 a  field  amplitude of $E_0 \cos \theta $, 
and a light  intensity of $\frac{1}{2}I_0 \cos^2 \theta $,
where  $E_0$ and $I_0$ are the amplitude and intensity of incident light,
respectively. Natural light is a completely unpolarized photon beam, 
which can be
decomposed into two independent orthogonal components of field. 
Similarly, the spin pair as a  whole is also completely unpolarized.
A photon beam passing through two polarizers is equivalent in effect 
to  a  spin-pair  beam  of which each spin passing through one 
Stern-Gerlach magnet,  therefore, they yield to the same cosine law of 
statistical joint outcome.

\par
Note that the wave function Eq. (7) is not new. It has been suggested 
soon after the EPR historical paper [3,4,12], but  finally it was 
refuted and abandoned due to two 
difficulties assumed. The first difficulty is that Eq. (7) do not 
conserve its form under  rotation [8], the second  difficulty
is that the  statistical joint outcome
is different from the standard one  shown in Eq. (8) (which is
the experimentally correct one) [13].
We now see that these two objections do  not stand in our model.  
The wave function Eq. (7) is rotationally invariant and yields the same 
 joint outcome [Eq. (8)] as the orthodox formulation, when the statistical
feature of the whole ensemble of spin pairs is concerned.  
Both models predict  definitely that 
two spins in each pair is always antiparallel to each other. 
If one is interested in the motion of single spin pair  
as to what direction the spin is along, then
neither our model nor the orthodox formulation can give  definite 
results, only statistical results can be predicted definitely.

\par
Several important differences can be  found between our model and
the orthodox formulation. 
 One  significant  consequence of Eq. (7) is that all
components of each spin can be measured simultaneously, since
now the two spins are noninteracting while in perfect correlation and thus
allow measuring one particle without affecting the other. 
This can be  accomplished by means of two spin-measuring 
 apparatuses (Stern-Gerlach magnets) oriented perpendicular
 to each other, say,
along the x-axis and y-axis, respectively.  Particles 1 and 2  passing
 through  apparatuses 1 and 2 will have their spin  components 
measured to be $S_{1x}$ and $S_{2y}$. Because of perfect antiparallel
correlation of the two spins, we at once know $S_{1y}=-S_{2y}$,
and  $S_{2x}=-S_{1x}$.  This is in contradiction with Heisenberg's
 uncertainty principle in the orthodox formulation, 
which asserts that  $S_x$ and $S_y$ of a particle
 can not be measured  simultaneously because they are non-commuting 
operators. 

\par
Another significant point of Eq. (7) is that
 the conventional concept of
wave-packet  reduction [5,6] is not present in our model,
because now each atom  is in a pure state with a definite spin vector. 
The measurement does not  determine  
what spin state the atom is in, but just tells 
us what it is in. From this viewpoint, the measurement is not an
 inseparable part  combined with the quantum system considered, 
a manner the orthodox  formulation strongly stresses. If one
wishes to uphold the concept of wave-packet reduction,
then this process must have happened during the process of
molecular dissociation. This also has  nothing to do with external
measurements by the observer. 

\par
In our model the perfect antiparallel correlation of the two spins is
maintained by the conservation law,
 while in the orthodox formulation
it is preserved by the nonlocal entanglement of spin wave function
 and realized in measurement by the mechanism of wave-packet reduction. 
In experiments  to test Bell's inequalities [17-20], two loopholes
 are generally supposed  to leave  the conclusion uncertain.
One is that only a small subset of all pairs are detected
due to  low detection/collection efficiency of the apparatus, the other
is that  the Einstein locality condition might not be maintained 
strictly.  These two loopholes are, however, not present in our model. 
The measurement process involved is always local,
and a small subset always represents fairly the whole ensemble
in a statistical manner.
It is not the communication between the two particles, but the
strict maintenance of conservation law in nature that upholds
the perfect correlation. From this point of view, our model
 agrees  with experimental observations in a more natural manner
than  the orthodox formulation does.

\par
Now we can see that our proposed model is realistic and local, 
and at the same time it is in as good  
agreement with experimental observations as the orthodox formulation,
when two-particle EPR pairs are concerned. 
From  the viewpoint of the orthodox formulation,  our model  
essentially lies inside the framework of hidden-variable theory. Then, since  
the hidden-variable theory constructed here  is local, realistic,  and 
makes same predictions as  orthodox quantum mechanics does, one
is led to conclude that Bell's theorem is not always true in regard to
two-particle EPR pairs.

\par 
Following the concept implied in above arguments, we turn to analyze the
scheme of entangled pair originally  proposed by EPR
[1]. The system they studied consists of two particles, and 
lies  in the state
$\Psi(x_1,x_2)=\delta(x_1-x_2-a)$, which is the eigenfunction of
the operator $x_1-x_2$ with eigenvalue $a$, and of the operator $p_1+p_2$
with eigenvalue 0. We emphasize 
 that this $\delta$-type wave function implies very strong
interaction between the two particles. This pair as a whole can not
be subject to a simultaneous measurement of the position and momentum 
without affecting each other. But the situation
changes completely   when the pair is dissociated  
at time $t_0$ and becomes  noninteracting by
a process that does not influence the total momentum.
Each particle is now  in a free state, with the momentum ($p_1$ and
$p_2$) specifying its motion feature (a pure state).
The conservation of total 
momentum satisfied at each event leads to a perfect correlation 
of the momentum  $p_1+p_2=0$,  from which 
we derive 
$$ x_1(t)- x_2(t)=a +p_1 (\frac{1}{m_1}+\frac{1}{m_2}) (t-t_0) \eqno(9) $$ 
under assumption of  free propagation of particles.
$m_1$ ($m_2$) is the mass of particle 1 (2), $t_0$ is the time when 
the particle dissociation occurs, and $t$ is the time at measurement. 

\par
Whether or not the position of the two particles persists a perfect correlation
 depends on the accuracy of determining at what time the dissociation of the
 molecule takes place.   If this can be accomplished, 
we immediately obtain a perfect correlation in both 
momentum and position, and  conclude that
both  the position and momentum of each  particle  can be measured 
simultaneously, and that  Heisenberg's uncertainty principle  
is violated.   Even if
we can not acquire a perfect correlation in  position, 
 we can still simultaneously 
measure the position and momentum of one of the two  particles. 
For example, we now decide to see  particle 1. We first measure 
its position and get a value
of $x_1$, at the  same time we measure the momentum of 
particle 2 and get a value
of $p_2$. Due to the perfect correlation of the  momentum, 
we at once know $p_1=-p_2$, and arrive at a position
that  the uncertainty principle is violated. 
Note that our argument is different from
original EPR's, where  the two particles are supposed to be always
 in a perfect entangled state (for both position and momentum)
when they are subject to measurements. In this entangled state, 
the momentum and position of a particle  can not be measured
simultaneously, because of the strong entanglement
 between the two composites.  We can further argue that 
in  Bohm's scheme  the violation of
uncertainty principle remains even if we only have a system with 
a perfect correlation in one of the spin components 
instead of all components.

\par
In the orthodox  quantum theory, it is assumed that
the wave function completely
specifies  the physical state of an individual system.
According to our local realistic model, it is possible to obtain 
through conservation laws in nature 
(which hold true in all mechanical theories) 
to obtain simultaneous knowledge
of non-commuting variables such as different components of a spin, and
the position and momentum of a particle. This  violates  
Heisenberg's uncertainty principle, one of the
basic foundations of orthodox quantum mechanics.  
Therefore, it is possible
to obtain more  complete information about individual system  
 beyond that allowed by  the  wave function  in the orthodox formulation, 
while at the same time arrive at the same statistical information
 on an ensemble of individual systems as that predicted  
by the orthodox formulation. 
This suggests that the wave function in orthodox quantum theory is
only a complete description of an ensemble of individual particles,
but not a complete description of individual particles.  
Let us look at the wave functions Eq. (5) and Eq. (7) once again.
They are equivalent to each other  when they are referred to an 
ensemble of EPR pairs, 
while  totally different when referred to an individual pair.
In the orthodox formulation, the wave function is always assumed to describe
an individual system. Now one finds  it is more likely that
the wave function [like Eq. (5)] only describes an ensemble of individual
systems.  It is  this difference in the wave function for
an individual system  that cause significant results of the proposed model
beyond the orthodox formulation.

\par
In summary,  we have proposed a local realistic model for quantum
mechanics of two-particle EPR pairs in the framework of standard quantum
mechanics for many-body systems. 
 In this model, the spooky nonlocal
characteristics present in the space-separated free EPR pairs 
as assumed in the orthodox formulation is abandoned. Also abandoned 
is the concept of wave-packet reduction  in the orthodox 
formulation  which is assumed to happen 
when external measurement occurs.   It is
the strict obedience of  conservation laws  in each  
event at the quantum level that maintain the perfect correlation
of two spatially-separated particles.
With a different starting viewpoint from orthodox quantum mechanics,  
the proposed model allows  more information
on individual particles beyond  the wave function in 
the orthodox formulation to be obtained.  
 At the same time it leads to the same statistical information on
an ensemble of individual particles as predicted by the orthodox 
formulation.
 It may be expected that the proposed local realistic model
 will help people  to  understand quantum mechanics in a conceptually 
 easy  way, as has been the case for classical physics.

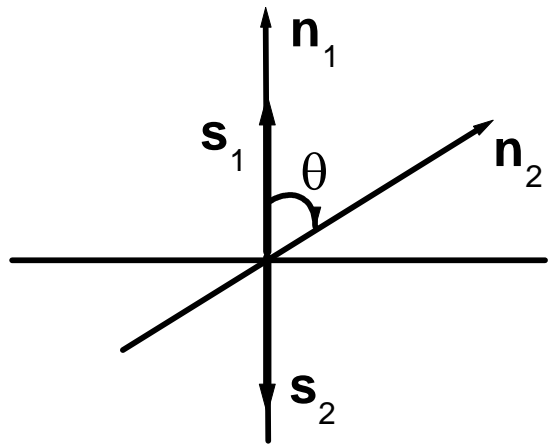
\begin{figure}
\caption{ Schematic configuration of an entangled spin-1/2 pair
with  antiparallel spins ${\bf s}_1$ and ${\bf s}_2$ 
 passed through two Stern-Gerlach
apparatuses along the ${\bf n}_1$ and ${\bf n}_2$ directions with an 
inclination of $\theta$. Due to rotational invariance and
completely unpolarized nature of the spin-pair beam, 
${\bf s}_1$ can be set to be parallel to ${\bf n}_1$.
}
\end{figure}

\end{document}